\documentclass[a4paper,fleqn]{cas-sc}

\usepackage[authoryear]{natbib}

\def\tsc#1{\csdef{#1}{\textsc{\lowercase{#1}}\xspace}}
\tsc{WGM}
\tsc{QE}
\tsc{EP}
\tsc{PMS}
\tsc{BEC}
\tsc{DE}

\begin{document}

\let\WriteBookmarks\relax
\def\floatpagepagefraction{1}
\def\textpagefraction{.001}
\shorttitle{Learning-Based Phase Estimation for CP Ranging}
\shortauthors{J. Bonczyk et~al.}

\title [mode = title]{Learning-Based Phase Estimation for Multi-Frequency Carrier Phase Ranging under Structured Multipath Conditions}

\author[1]{Jakub Bonczyk}[
]
\cormark[1]
\ead{jakub.bonczyk@put.poznan.pl}

\author[1]{Jakub Nikonowicz}[
orcid=0000-0001-9770-1177
]
\cormark[2]
\ead{jakub.nikonowicz@put.poznan.pl}

\author[1]{{\L}ukasz Matuszewski}[
orcid=0000-0002-5795-3246
]
\ead{lukasz.matuszewski@put.poznan.pl}

\affiliation[1]{
organization={Faculty of Computing and Telecommunications, Poznan University of Technology},
city={Poznan},
country={Poland}
}

\cortext[cor1]{Corresponding author}
\cortext[cor2]{Principal corresponding author}

\nonumnote{This work was supported by the Polish Ministry of Science and Higher Education.}

\begin{abstract}
Carrier-phase (CP) ranging is a key enabler of high-precision positioning in modern wireless systems. In multi-frequency OFDM-based sensing, phase observations across subcarriers provide information about the underlying propagation geometry. However, in realistic industrial and urban environments, these observations exhibit non-Gaussian and asymmetric characteristics due to deterministic multipath components, violating standard circular statistical assumptions.
In this work, we analyze CP-based ranging as an estimation problem over circular phase observations. We show that conventional model-based estimators, such as circular averaging under von Mises assumptions, become biased under 3GPP-compliant propagation conditions. Using a QuaDRiGa-based simulation framework, we evaluate empirical phase distributions in Industrial Factory (InF) and Urban Microcell (UMi) scenarios and quantify their deviation from classical statistical models.
To address these limitations, we propose a learning-based estimator that operates directly on empirical phase distributions without assuming a predefined statistical model. Experimental results show improved accuracy compared to classical estimators, particularly under multipath conditions.
\end{abstract}




\maketitle

\makeatletter
\renewenvironment{figure}[1][]%
  {\par\vspace{0.8em}\noindent\begin{minipage}{\linewidth}\def\@captype{figure}\centering}%
  {\end{minipage}\par\vspace{0.8em}}

\renewenvironment{table}[1][]%
  {\par\vspace{0.8em}\noindent\begin{minipage}{\linewidth}\def\@captype{table}\centering}%
  {\end{minipage}\par\vspace{0.8em}}
\makeatother

\section{Introduction}

High-precision positioning is a key feature in the development of 5G-Advanced and 6G systems. Recent standardization efforts in 3GPP, spanning Releases 18 through 20, include carrier-phase (CP) techniques to improve localization accuracy across diverse environments \citep{R1-2306873,3GPP_TR21_918,3GPP_TR38_859,cha2025}. While traditional methods rely on time-based and angle-based estimations, CP positioning exploits phase observations across multiple subcarriers to achieve higher resolution, provided that a sufficient Signal-to-Noise Ratio (SNR) is maintained \citep{cha2025,fouda2022,nikonowicz2024}.

The implementation of CP-based ranging in modern systems leverages the frequency-domain structure of OFDM waveforms. The deterministic spacing between subcarriers enables the extraction of phase information across the signal bandwidth, allowing phase differences between subcarriers to be exploited to synthesize a \textit{virtual wave} \citep{fouda2022,nikonowicz2024,talebian2026}. By configuring the Subcarrier Spacing (SCS) together with the signal comb structure---where specific subcarriers are allocated for positioning---the resulting virtual wavelength can be adjusted to match the operational distance range, e.g., the Inter-Site Distance (ISD). This approach is closely related to the Integer Ambiguity (IA) problem: since phase observations are restricted to the $[0, 2\pi)$ interval, the total number of full wavelengths remains unknown. Selecting the virtual wavelength on the order of the expected link distance avoids ambiguity and simplifies the estimation process, eliminating the need for complex ambiguity resolution algorithms \citep{talebian2026}.

These concepts are reflected in current 3GPP standardization efforts. CP-based methods are supported for both uplink and downlink through Sounding Reference Signals (SRS) and Positioning Reference Signals (PRS). Notably, starting from Release 18, 3GPP specifications define the reporting of Reference Signal Carrier Phase (RSCP) \citep{cha2025}. As the standard does not mandate a specific frequency reference for RSCP, it can be effectively used to report the phase (e.g., aggregated) of a virtual wave derived from multiple frequency-pair measurements across the signal bandwidth. Similar approaches can be applied to other systems, such as IEEE 802.11az (and the recently ratified 802.11bk). While current IEEE-based Fine Timing Measurements (FTM) utilize bandwidths up to 320~MHz \citep{kosekszott2026}, carrier-phase extensions represent a promising direction for further improving localization precision.

However, CP-based ranging in real-world environments such as Industrial Factory (InF) and Urban Microcell (UMi) scenarios is affected by propagation conditions that differ  from sparse-clutter, open-environment models \citep{abuyaghi2025,talebian2024,talebian2026}. Standard approaches often assume that phase noise follows a von Mises distribution \citep{lan2026,matthe2023,morales2026,wymeersch2023} or its mixtures \citep{cheng2024}, while wrapped Cauchy distributions have been considered for heavy-tailed cases \citep{godana2013}. These assumptions, however, frequently fail in environments characterized by deterministic reflections and multipath clusters, as defined in 3GPP TR 38.901 \citep{3GPP_TR38_901}. As a result, the observed phase distributions are strongly shaped by the propagation environment and cannot be reliably captured by simple parametric models.

In this paper, we analyze the phase distribution for CP-based ranging from a single transmitter in multi-subcarrier systems. We simulate UMi and InF environments using the QuaDRiGa channel generator, which strictly follows 3GPP modeling standards. Based on the simulated empirical phase distributions, we show that classical estimators lead to biased phase estimates under realistic propagation conditions. To address this limitation, we propose learning-based estimators—specifically Fully Connected (FC), 1D Convolutional Neural Networks (1D CNN) and 2D Convolutional Neural Networks (2D CNN)—that operate directly on empirical phase distributions. By leveraging the structure present in the observations, the proposed approach enables robust phase estimation. In this formulation, the estimation problem is treated as a data-driven inference task, where the underlying structure of the observations is learned rather than explicitly modeled.


\textcolor{black}{The main contributions of this work are threefold. First, we demonstrate that empirical phase distributions under standardized 3GPP propagation models deviate substantially from conventional circular statistical assumptions, leading to systematic bias in classical model-based estimators. Second, we introduce a configuration-independent histogram representation together with learning-based estimators that operate directly on empirical phase distributions without assuming a predefined statistical model. Third, we provide a comprehensive evaluation under multiple bandwidths, SNR regimes, receiver mobility conditions, and PRS aggregation strategies, comparing the proposed approach against conventional circular averaging and model-based distribution fitting methods.}

\textcolor{black}{The remainder of this paper is organized as follows. Section~\ref{sec:vw} introduces the virtual wavelength framework and positions it with respect to existing carrier-phase ranging methods. Section~\ref{sec:phase} reviews the statistical estimation approaches and discusses their limitations under structured multipath propagation. Section~\ref{sec:sim} describes the simulation framework. Section~\ref{sec:nn} introduces the proposed learning-based estimator and discusses its relationship to existing data-driven approaches. Section~\ref{sec:exp} presents the experimental evaluation, followed by the conclusions in Section~\ref{sec:con}.}

\section{Virtual Wavelength Framework in Multi-Carrier Phase Ranging}
\label{sec:vw}

Carrier-phase measurements enable high-resolution estimation of the propagation distance $d$ between the transmitter (Tx) and receiver (Rx). However, for a single carrier frequency, the observed phase $\phi \in [0, 2\pi)$ provides only the phase remainder modulo $2\pi$, corresponding to the last incomplete cycle of propagation. In multi-carrier systems, this limitation is addressed by exploiting phase observations across different subcarriers. Consider two subcarriers at frequencies $f_1$ and $f_2$. After removing the known pilot symbols, the phase observations can be written as~\citep{talebian2026}

\begin{equation}
    \phi_1 = \frac{2\pi d}{\lambda_1} + \theta + 2\pi N_1,
\end{equation}
\begin{equation}
    \phi_2 = \frac{2\pi d}{\lambda_2} + \theta + 2\pi N_2,
\end{equation}

\noindent where $\phi_1,\phi_2 \in [0,2\pi)$ denote the wrapped phase observations, $\theta$ is the common phase offset between Tx and Rx oscillators, and $N_1,N_2 \in \mathbb{N}_0$ denote the number of complete carrier cycles accumulated over the propagation path.

By forming the phase difference $\Delta \phi = \phi_2 - \phi_1$, the common offset $\theta$ is eliminated, yielding

\begin{equation}
    \Delta \phi = 2\pi d \left( \frac{f_2 - f_1}{c} \right) + 2\pi \Delta N,
\end{equation}

\noindent where $\Delta N = N_2 - N_1$. This expression can be rewritten as

\begin{equation}
    \Delta \phi = \frac{2\pi d}{\lambda_v} + 2\pi \Delta N,
\end{equation}

\noindent where the virtual wavelength is defined as

\begin{equation}
    \lambda_v = \frac{c}{\Delta f}, \quad \Delta f = f_2 - f_1.
    \label{eq:lambda_v}
\end{equation}

\noindent The phase difference therefore corresponds to a synthetic wave with wavelength $\lambda_v$, which is typically much larger than the individual carrier wavelengths, as it depends inversely on the frequency difference $\Delta f$, as given in (\ref{eq:lambda_v}). This operation simultaneously removes the oscillator offset and extends the effective wavelength, increasing the unambiguous observation range.

At this point, the integer ambiguity appears in the form of $\Delta N$, leading to the distance expression \citep{R1-2306873,talebian2026}

\begin{equation}
    d = \left( \frac{\Delta \phi}{2\pi} + \Delta N \right)\lambda_v.
\end{equation}

\noindent The ambiguity is now defined with respect to the virtual wavelength rather than the carrier wavelength. Consequently, if the propagation distance satisfies $d < \lambda_v$, then $\Delta N = 0$ and the solution becomes unique without requiring explicit ambiguity resolution.

In OFDM-based systems, the frequency difference $\Delta f$ is determined by the subcarrier spacing $\Delta f_{\text{SCS}}$ and the index separation $k$ between subcarriers, i.e., $\Delta f = k\,\Delta f_{\text{SCS}}$. Since $k$ is limited by the number of active subcarriers, the maximum practical separation is determined by the occupied bandwidth rather than by the nominal channel bandwidth. This allows the virtual wavelength to be scaled over several orders of magnitude within a given OFDM allocation. For adjacent subcarriers ($k=1$), an SCS of $30$~kHz in FR1 gives $\lambda_v \approx 10$~km, while an SCS of $120$~kHz in FR2 gives $\lambda_v \approx 2.5$~km. At the opposite extreme, for an allocation of 3276 active subcarriers, selecting subcarriers separated by half of the active allocation gives $k=1638$. This corresponds to $\Delta f = 49.14$~MHz and $\lambda_v \approx 6.10$~m for FR1 with $30$~kHz SCS, and to $\Delta f = 196.56$~MHz and $\lambda_v \approx 1.53$~m for FR2 with $120$~kHz SCS. These two cases illustrate the scalability of the virtual wavelength over several orders of magnitude. Large $\lambda_v$ ensures ambiguity-free ranging over typical link distances, while small $\lambda_v$ provides high sensitivity of the phase-to-distance mapping. This trade-off enables flexible system design, where the virtual wavelength can be matched to the expected propagation range and refined using additional frequency separations within the available bandwidth.
At the same time, the achievable ranging accuracy remains fundamentally limited by the phase measurement error, which scales proportionally with $\lambda_v$. Consequently, large virtual wavelengths improve robustness to ambiguity at the cost of reduced precision, whereas smaller $\lambda_v$ enhance resolution but require ambiguity handling \citep{fan2021}.

Importantly, in OFDM-based systems all subcarriers are observed within a single symbol, allowing multiple frequency separations $\Delta f$ to be formed from a single measurement. This enables the simultaneous construction of multiple virtual wavelengths, which can be jointly exploited to first eliminate the ambiguity using large $\lambda_v$ and subsequently refine the distance estimate using smaller $\lambda_v$, without requiring multi-epoch processing. This multi-scale observation capability is a direct consequence of the wideband OFDM structure and distinguishes it from single-carrier phase measurements \citep{li2022,3GPP_R1_2306873}.

\begin{figure}[!htb]
    \centering
    \includegraphics[width=0.65\linewidth]{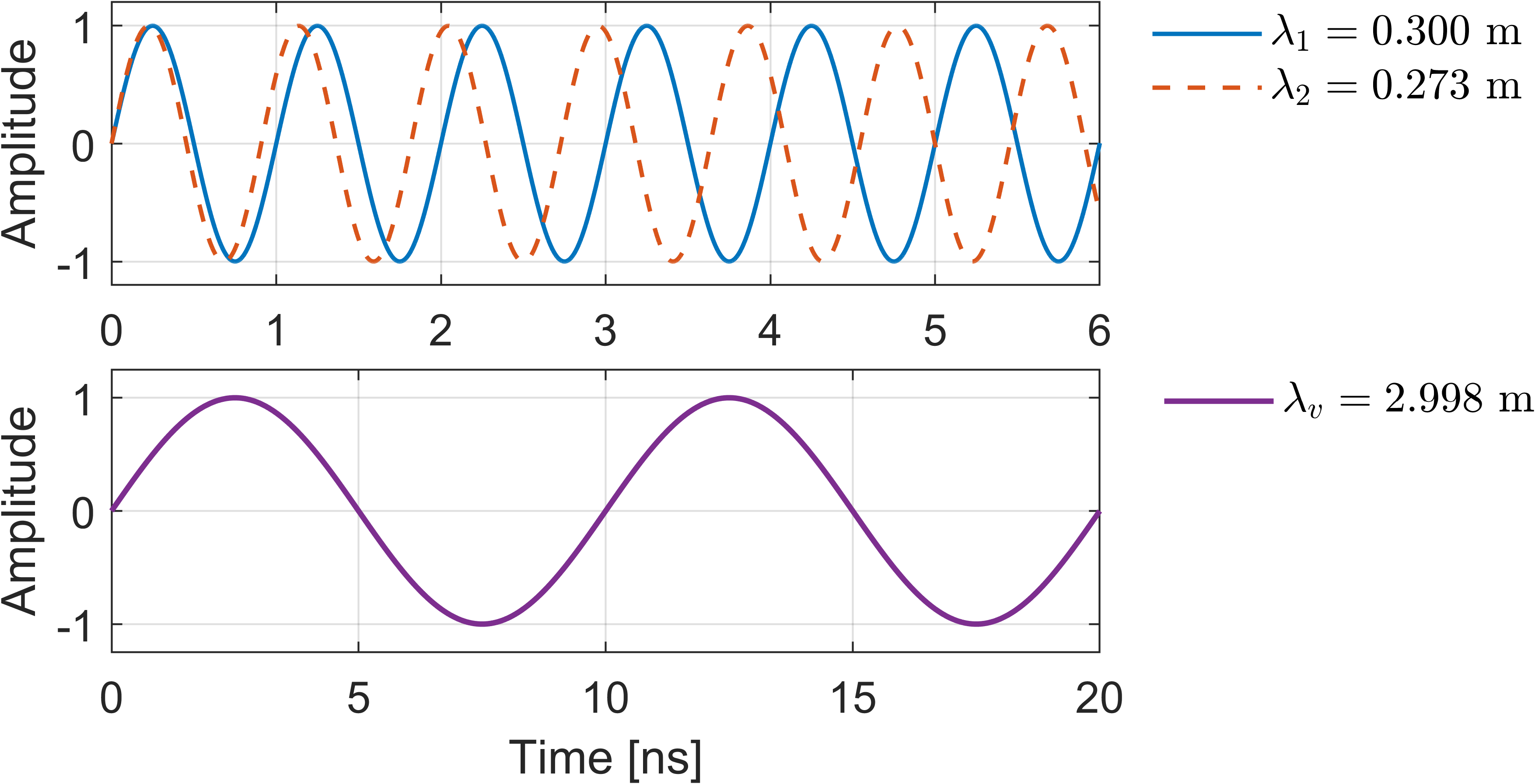} 
    \caption{Conceptual representation of virtual wave $\lambda_v$ generation from the phase difference $\Delta\phi$ of two subcarriers $\lambda_1$, $\lambda_2$.}
    \label{fig:virtual_wave}
\end{figure}

\section{Stochastic and Deterministic Deviations in Phase Distributions}
\label{sec:phase}

The performance of conventional CP-based ranging is fundamentally tied to the statistical properties of phase observations aggregated across the virtual wavelength framework. Standard estimators typically rely on circular averaging, often weighted by amplitude, to mitigate noise effects \citep{li2022,cho2025,3GPP_R1_2306873}. Such approaches implicitly assume that phase errors are independent and symmetrically distributed, commonly modeled by a von Mises distribution centered at the true phase of arrival. Under this assumption, low-amplitude components, including deep fades, are either averaged out or effectively suppressed due to their reduced amplitude contribution. However, this assumption is not consistent with the propagation models used in environments with cluster-based propagation, such as InF and UMi scenarios. In fact, the presence of a limited number of dominant clusters is an inherent property of the standardized 3GPP TR 38.901 channel models \citep{3GPP_TR38_901}.

\begin{figure}[!htb]
    \centering
    \includegraphics[width=0.75\linewidth]{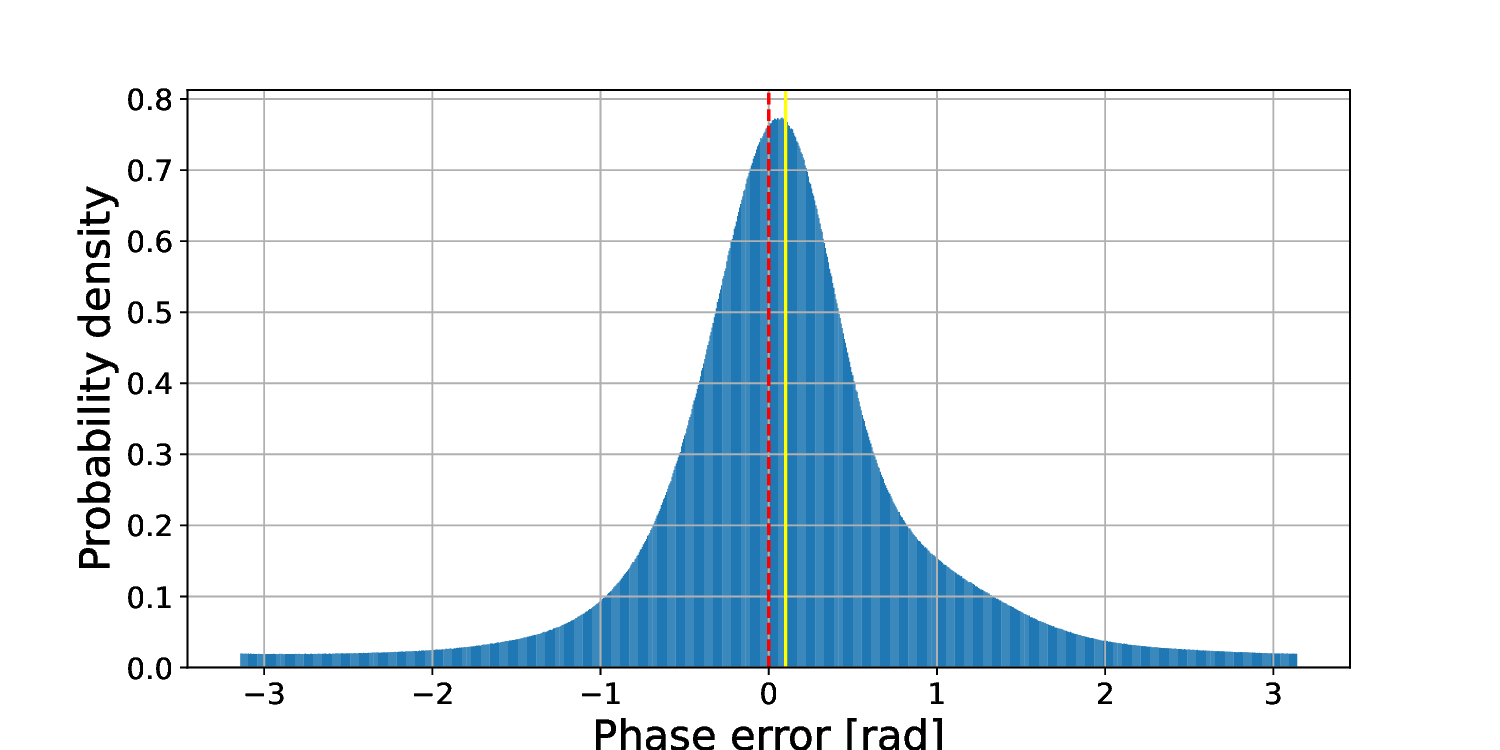} 
    \caption{Empirical phase error distribution in the InF LOS scenario, exhibiting multimodal characteristics. The red dashed line denotes zero phase error, while the yellow line corresponds to the circular mean (0.10 rad).}
    \label{fig:phase_analysis_inf}
\end{figure}

\begin{figure}[!htb]
    \centering
    \includegraphics[width=0.75\linewidth]{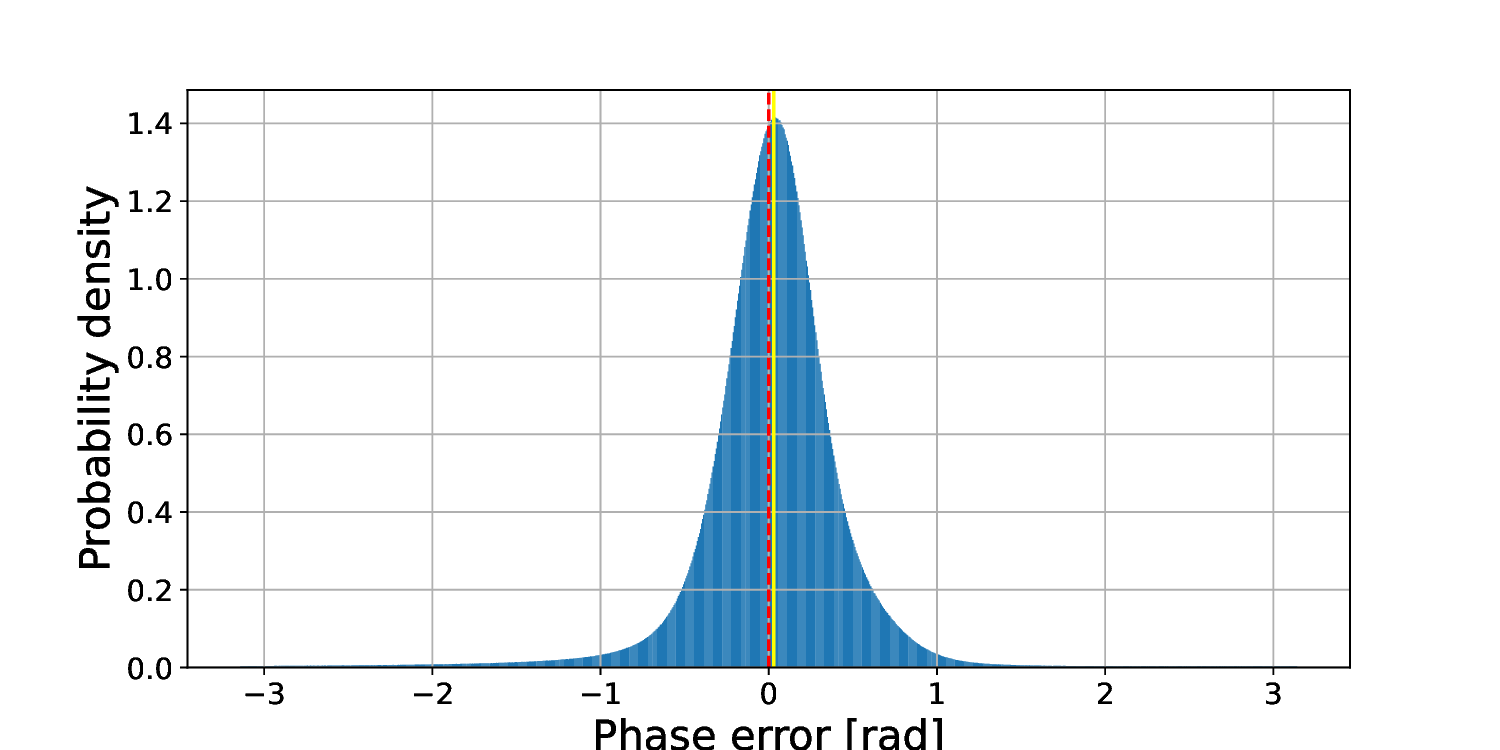} 
    \caption{Empirical phase error distribution in the UMi LOS scenario, exhibiting multimodal characteristics. The red dashed line denotes zero phase error, while the yellow line corresponds to the circular mean (0.03 rad).}
    \label{fig:phase_analysis_umi}
\end{figure}

According to the standardized channel models, propagation is characterized by a limited number of dominant clusters corresponding to specular reflections from large objects such as metallic surfaces, industrial equipment, and urban infrastructure \citep{3GPP_TR38_901}. These components introduce deterministic phase shifts that remain consistent across groups of subcarriers while maintaining significant signal energy. As a result, they do not average out but instead introduce a systematic bias in the estimated phase. Consequently, the resulting phase distribution deviates from the assumed unimodal and symmetric form, exhibiting asymmetry, multimodality, and heavy-tailed behavior. This behavior is therefore not a consequence of measurement imperfections, but a direct result of the underlying propagation structure imposed by the channel model, as illustrated in Fig.~\ref{fig:phase_analysis_inf} and Fig.~\ref{fig:phase_analysis_umi}.

\begin{figure}[!htbp]
    \centering
    \includegraphics[width=0.75\linewidth]{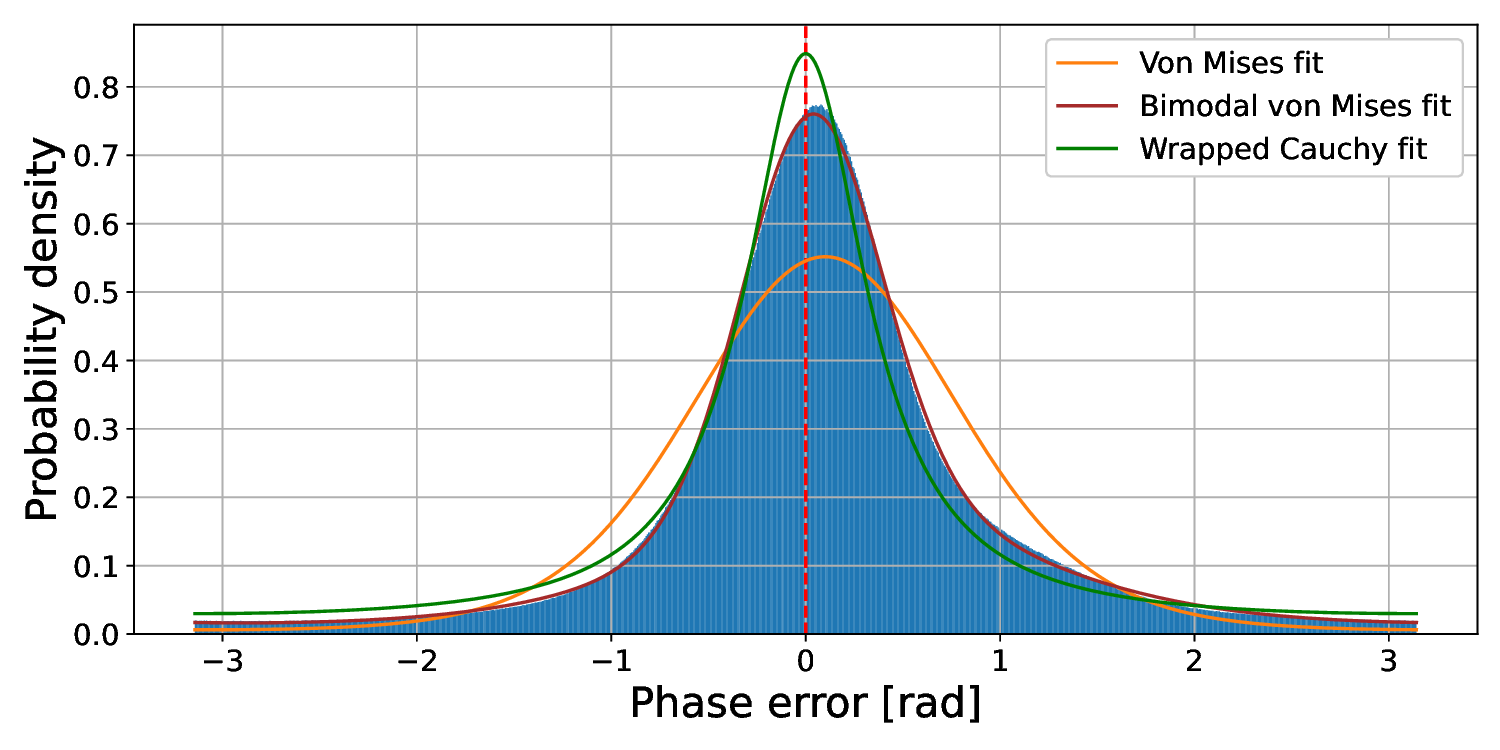}
    \caption{Comparison of empirical phase error data with theoretical distribution fits for the InF LOS scenario.}
    \label{fig:phase_analysis_inf_aic}
\end{figure}

\begin{figure}[!htbp]
    \centering
    \includegraphics[width=0.75\linewidth]{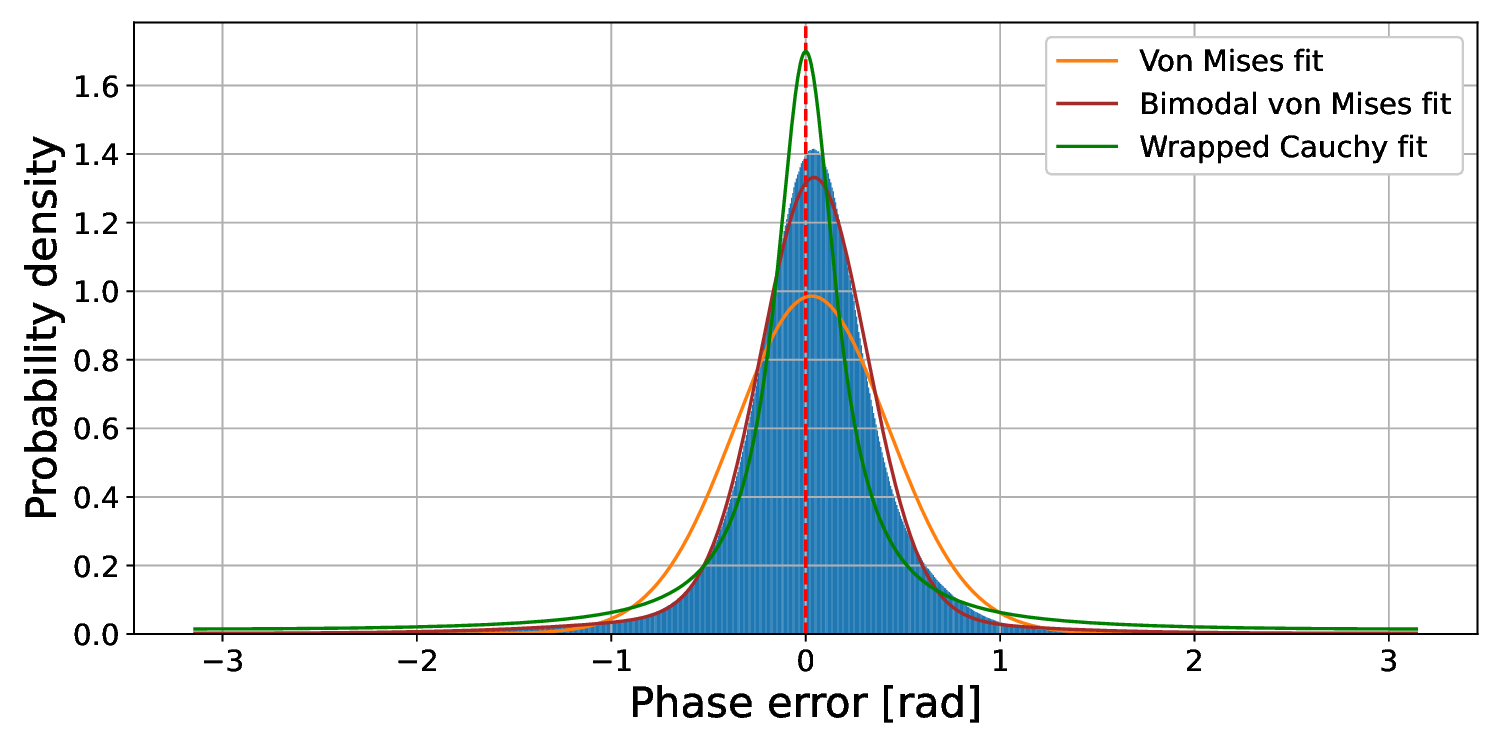}
    \caption{Comparison of empirical phase error data with theoretical distribution fits for the UMi LOS scenario.}
    \label{fig:phase_analysis_umi_aic}
\end{figure}

To quantify this mismatch, candidate models were evaluated using the Akaike Information Criterion (AIC) for three reference distributions: the von Mises distribution, the wrapped Cauchy distribution, and a bimodal von Mises mixture. AIC provides a relative measure of model quality by balancing goodness-of-fit with model complexity. The results, summarized in Table~\ref{tab:aic_tab}, show that although the bimodal von Mises model achieves the lowest AIC value, it still fails to accurately capture the tail behavior and the shift of the dominant mode observed in the empirical data, as illustrated in Fig.~\ref{fig:phase_analysis_inf_aic} and Fig.~\ref{fig:phase_analysis_umi_aic}. Moreover, the improved fit comes at the cost of increased model complexity, without a proportional improvement in capturing the underlying distribution characteristics.

\begin{table}[!htbp]
\caption{Akaike Information Criterion (AIC) for phase distribution models}
    \centering
    \begin{tabular}{l|c}
    \hline
        Distribution Model & AIC Value \\
        \hline
        InF LOS & \\
        
        von Mises  & 175959452.127 \\
        wrapped Cauchy  & 170993875.476 \\
        bimodal von Mises  & 168205703.704 \\
        \hline

        \hline
        UMI LOS & \\
        
        von Mises  & 66663832.423 \\
        wrapped Cauchy  & 60491668.547 \\
        bimodal von Mises  & 52400409.917 \\
        \hline
        
    \end{tabular}
    \label{tab:aic_tab}
\end{table}

\textcolor{black}{Under practical sample sizes, the asymmetry and deviation from classical models become even more pronounced. As a consequence, circular averaging yields biased estimates, as dominant deterministic components displace the distribution from the true phase of arrival. This effect cannot be mitigated by amplitude weighting, since reflected components retain significant energy. These patterns are not purely random but are largely induced by the underlying propagation geometry and the local multipath environment encountered along the receiver trajectory, indicating the limitations of model-based statistical estimators. Rather than being treated as random perturbations, these structured phase distributions can instead be regarded as informative representations of the underlying propagation conditions. This observation motivates the learning-based estimation approach introduced in Section~\ref{sec:nn}.}


\section{Simulation Framework and Signal Configuration}
\label{sec:sim}

To evaluate the performance of the proposed estimators under the non-ideal phase distributions established in Section~\ref{sec:phase}, a simulation framework based on the Quasi-Deterministic Radio Channel Generator (QuaDRiGa) is employed \citep{quadriga}. The generator is configured according to the 3GPP TR~38.901 channel model, ensuring consistency with cluster-based propagation, delay spreads, angular characteristics, and frequency selectivity for InF and UMi scenarios. All simulations are conducted under LOS conditions, with multipath components inherently included by the standardized channel model.

The considered setup corresponds to a single-link ranging scenario between a stationary Tx and a mobile Rx. The Tx position is fixed, while the initial Rx position is randomly drawn from a uniform distribution over a two-dimensional deployment area defined by the ISD. The Rx follows a linear trajectory of length $16.65$~m with a constant velocity of $v = 30$~km/h. \textcolor{black}{The Rx depending on the mobility configuration follows a linear trajectory of maximum length $16.65$~m with a constant velocity of $v = 30$~km/h or a linear trajectory of maximum length $1.665$~m with a constant velocity of $v = 3$~km/h. Trajectory lengths are derived from a duration of single simulation trial.} The trajectory direction is randomly selected \textcolor{black}{from a uniform distribution} under the constraint that the entire trajectory remains within the considered deployment area, avoiding boundary effects related to the ISD. This setup introduces a controlled temporal evolution of the channel, including Doppler effects and gradual decorrelation of multipath components. For each realization, a new clutter configuration is generated, ensuring variability of the propagation environment. Both Tx and Rx are equipped with omnidirectional antennas. The Tx antenna height is set to $8$~m for the InF scenario and $10$~m for the UMi scenario, while the Rx antenna height is fixed at $1.5$~m.

The transmitted signal follows a 5G NR-compliant PRS configuration. The carrier frequency is set to 3.5~GHz, and the system operates with a total bandwidth of 100~MHz and an SCS of 30~kHz, resulting in an FFT size of 4096. Out of these, 3276 subcarriers are active, and a Comb-2 pattern is applied, meaning that every second subcarrier is used for PRS transmission, yielding 1638 subcarriers per OFDM symbol employed for phase-based ranging.\textcolor{black}{The system was also tested with a second configuration in which the total bandwidth was limited to 50~MHz while keeping SCS of 30~kHz. FFT size remained the same. However, only 1596 subcarriers are active. Since Comb-2 pattern is applied, only 798 per symbol are employed for ranging}. PRS symbols are transmitted over 12 out of 14 OFDM symbols within each slot and in every slot of the radio frame, resulting in a high-density temporal configuration that enables controlled aggregation of phase observations without introducing scheduling sparsity. To model realistic reception conditions, \textcolor{black}{AWGN is applied to the complex baseband channel observations, with the SNR maintained within the two ranges of 3--6~dB and 10--12~dB.} The complete set of system parameters is summarized in Table~\ref{tab:params}.

\textcolor{black}{The empirical phase distributions analyzed in Section~\ref{sec:phase} were obtained by aggregating a large number of phase samples (38400 PRS symbols for UMi and 48000 for InF, with 1613 and 1538 phase differences per symbol, respectively). Such large-scale aggregation was used solely to characterize the underlying phase-error statistics. In practical operation, however, phase observations are typically available from only one or a small number of consecutive PRS symbols, resulting in substantially fewer samples for estimation.}

\begin{table}[!htb]
\caption{System and Simulation Parameters}
\label{tab:params}
\centering
\begin{tabular}{lc}
\hline
Parameter & Value \\
\hline
\multicolumn{2}{l}{\textit{Signal Configuration}} \\
Carrier Frequency & 3.5 GHz \\
Total Bandwidth & 100 MHz \textcolor{black}{, 50MHz} \\
Subcarrier Spacing (SCS) & 30 kHz \\
FFT Size & 4096 \\
Active Subcarriers & 3276 \textcolor{black}{, 1596} \\
PRS Subcarriers & 1638 (Comb-2) \textcolor{black}{, 798} \\
PRS Symbols per Slot & 12 (of 14) \\
\hline
\multicolumn{2}{l}{\textit{Channel Model}} \\
3GPP Scenario (InF) & 38.901\_InF\_LOS \\
3GPP Scenario (UMi) & 38.901\_UMi\_LOS \\
Additional Noise & AWGN (post-channel) \\
SNR & 10--12 dB \textcolor{black}{, 3--6 dB} \\
\hline
\multicolumn{2}{l}{\textit{Deployment and Mobility}} \\
Antenna Height Tx (InF) & $8$ m \\
Antenna Height Tx (UMi) & $10$ m \\
Antenna Height Rx & $1.5$ m \\
Inter-Site Distance (InF) & 50 m \\
Inter-Site Distance (UMi) & 200 m \\
Trajectory Length & $16.65$ m \textcolor{black}{, $1.66$ m} \\
Rx Velocity & 30 km/h \textcolor{black}{, 3 km/h} \\
\hline
\multicolumn{2}{l}{\textit{Derived Configuration}} \\
Virtual Wavelength (InF) & $\approx 50$ m \\
Virtual Wavelength (UMi) & $\approx 200$ m \\
\hline
\end{tabular}
\end{table}

The ranging process operates on differential phase observations formed between subcarriers, as described in Section~\ref{sec:vw}. The effective virtual wavelength $\lambda_v$ is determined by the selected subcarrier separation $\Delta f$ and is chosen to be on the order of the ISD, ensuring ambiguity-free estimation ($\Delta N = 0$) for the considered deployment scenarios. In particular, this corresponds to $\lambda_v \approx 50$~m for the InF scenario and $\lambda_v \approx 200$~m for the UMi scenario.

\textcolor{black}{Phase-based observations are aggregated over consecutive PRS symbols, considering configurations with $\{1, 12, 60\}$ symbols. Additionally, phase-based observations are derived from one PRS symbol per slot, with temporal aggregation performed over $\{5,10\}$ consecutive slots, capturing more diverse channel response while maintaining smaller size.} This enables the analysis of the trade-off between improved statistical representation and motion-induced phase decorrelation. For a receiver moving with velocity $v$, the accumulated displacement over the observation interval remains small compared to the virtual wavelength, i.e., $\Delta d \ll \lambda_v$. Under this condition, the phase evolution induced by motion does not significantly distort the periodic structure of the virtual wave, and the aggregated measurements can be interpreted as originating from a quasi-static geometry.  At the same time, increasing the number of measurements improves the robustness of the empirical phase distribution but also extends the observation interval, thereby increasing the displacement. As a result, this condition may eventually be violated, leading to phase inconsistency.

\textcolor{black}{To quantify the correlation between PRS symbols, autocorrelation was computed from the received complex symbol observations after removing the known PRS sequence. The observations from each simulation trial were divided into non-overlapping windows of 120 consecutive PRS symbols. For each window, autocorrelation was computed as a function of the lag expressed in PRS symbols. All results within the window were averaged by the number of active subcarriers and then normalized at zero lag. Resulting autocorrelation values were averaged across all windows. This procedure was repeated for one trial for each channel model and Rx velocity, resulting in 4 separate results.}

\begin{figure}
    \centering
    \includegraphics[width=1\linewidth]{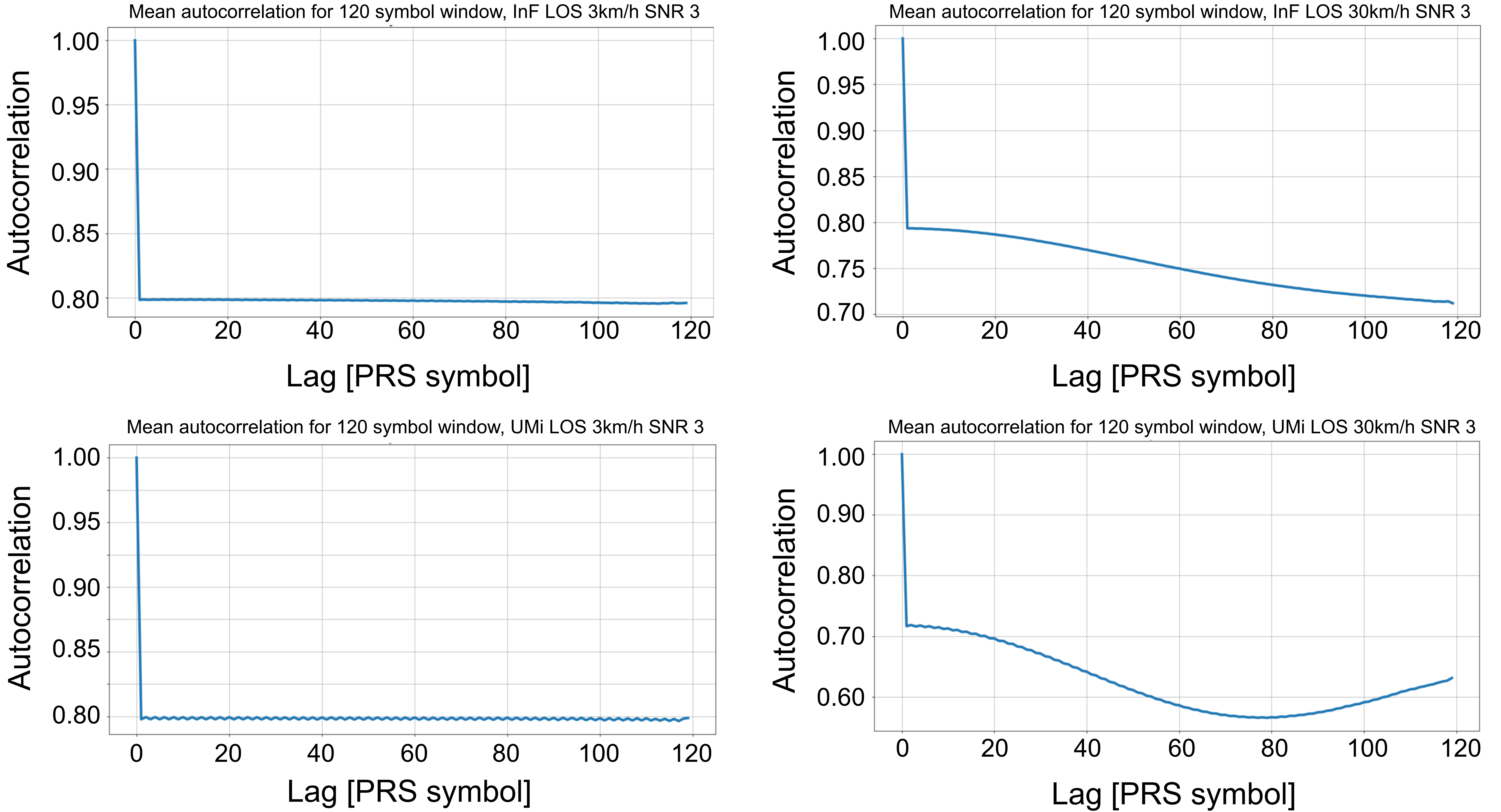}
    \caption{Comparison of autocorrelation between multiple scenarios}
    \label{fig:autocorr}
\end{figure}

\textcolor{black}{Results in Fig.~\ref{fig:autocorr} show that the PRS symbols remain strongly correlated over the span of 120 PRS symbols. At 3 km/h, correlation values remain stable across the whole window, while correlation at an Rx speed of 30 km/h begins to decrease gradually after approximately 20 PRS symbols. These observations provide a physical explanation for the performance differences obtained with different PRS aggregation lengths, as discussed in Section~\ref{sec:exp}.}

\section{Data-Driven Phase Estimation}
\label{sec:nn}

Given the limitations of statistical averaging, we propose a data-driven approach that directly estimates the true phase from the empirical distribution. The estimator learns a mapping from observed phase patterns to the underlying propagation delay, enabling adaptation to propagation-dependent patterns present in the data.

\begin{figure}[!htbp]
    \centering
    \includegraphics[width=0.65\linewidth]{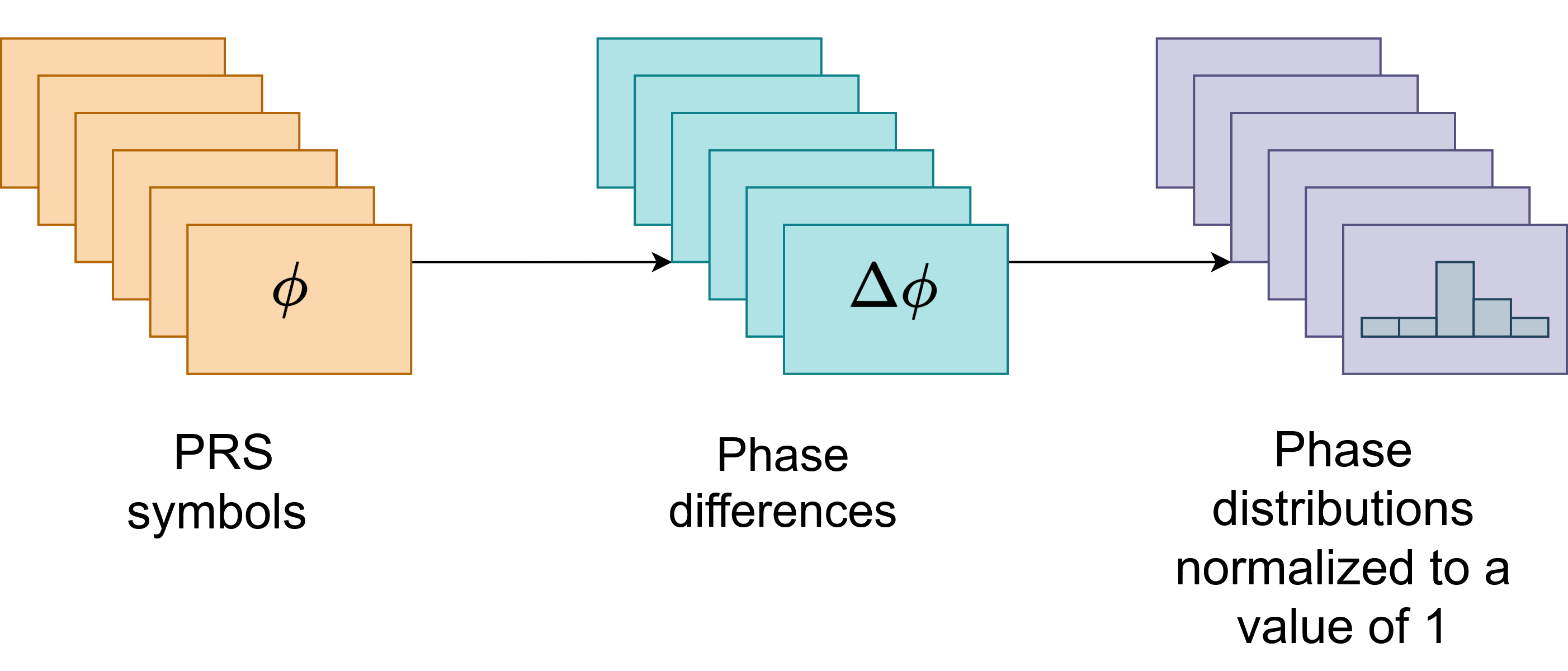} 
    \caption{Data processing pipeline from PRS signal acquisition.}
    \label{fig:pipeline}
\end{figure}

The model input comprises phase differences extracted from subcarrier pairs across multiple PRS symbols. These measurements form a structured representation of the phase distribution over frequency and time. Each PRS symbol in a single sample is represented as a phase distribution with a phase resolution of 0.01 radians over the interval $[0,2\pi)$. Each distribution is also normalized to 1 (Fig.~\ref{fig:pipeline}).

\textcolor{black}{Such representation allows the estimator to incorporate phase measurements derived from different PRS configurations. Since the phase histogram is normalized before estimation, the absolute number of phase-difference observations is not retained. Consequently, the input does not explicitly encode the number of allocated resource blocks, the comb size, or the total number of utilized subcarriers. Differences between PRS configurations are instead reflected implicitly through the shape of the empirical phase distribution. As a result, the proposed representation provides a fixed-dimensional input regardless of the underlying PRS configuration, enabling the same neural estimator to be applied across different signal bandwidths, subcarrier spacings, and PRS resource allocations without modifying the network architecture.}

\textcolor{black}{In contrast to learning-based approaches that operate directly on raw phase observations, e.g., \cite{Ayten2025, cheng2026, Long2026}, whose input dimensionality depends on the measurement configuration, the proposed histogram-based representation decouples the estimator from the physical signal parametrization while preserving the statistical information required for accurate phase estimation.}

\begin{figure}[!htbp]
    \centering
    \includegraphics[width=0.85\linewidth]{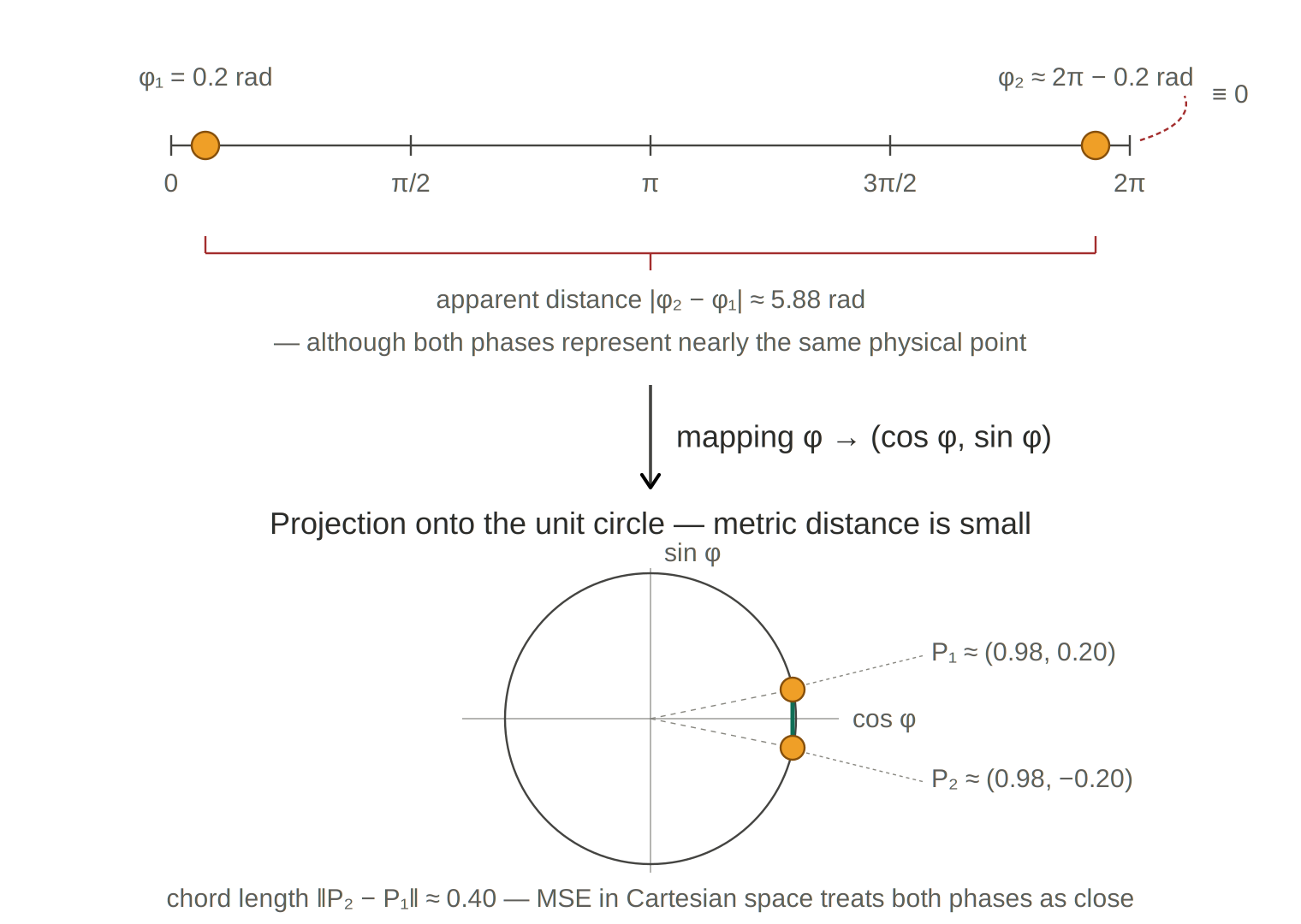}
    \caption{The phase-discontinuity problem on the linear axis (top) and its resolution via the projection onto the unit circle (bottom).}
    \label{fig:phase_wrap}
\end{figure}

In training deep models on cyclic data, directly regressing the phase value $\phi$ in the interval $[0, 2\pi)$ is problematic because the endpoints $0$ and $2\pi$ represent the same point in phase space (Fig.~\ref{fig:phase_wrap}). To eliminate this problem, the proposed method uses a phase projection onto a two-dimensional feature vector $[\cos\phi,\, \sin\phi]$, which maps the data directly onto the unit circle \citep{Pavllo2020}.
Using the mean squared error (MSE) on Cartesian components offers several advantages over the cosine similarity loss function. First, MSE on normalized vectors yields a more stable, linear gradient in the final optimization phase, which is crucial for accurately aligning the estimate with the reference value. Second, combining a unit normalization layer with the MSE function forces the network to generate results strictly on the circumference of the circle, preventing numerical instabilities and trivial solutions (such as vector collapse to zero) that can occur with pure angle minimization. As a result, the model learns nonlinear mappings between the observed distribution and the actual signal delay more efficiently, leading to greater training stability and more accurate prediction of the final phase.
During inference, the predicted phase is recovered using the 2-argument arctangent function.

The ground-truth phase is derived from the known Tx and Rx positions, enabling supervised training. The objective function minimizes circular error between predicted and reference phase values.

\textcolor{black}{Three neural architectures are considered}. The FC model captures global dependencies across the entire bandwidth, providing a low-complexity baseline. The 1D CNN enables detection of consistent phase distortions across adjacent histogram bins, while using multiple histograms from sample as input channels. \textcolor{black}{The 2D CNN exploits both distortions within the PRS symbol dimension and histogram bins.} Both of those architectures are trained using the Adam optimizer with a learning rate of 0.0001 with mean squared error as a loss function. \textcolor{black}{All architectures are trained using the Adam optimizer with a learning rate of 0.0001 with mean squared error as a loss function.} Training is performed for a maximum of 50 epochs. During training, model performance is evaluated on the validation dataset. Early stopping using validation loss is applied to prevent overfitting, restoring the model weights to values that correspond to the lowest validation loss.

To prevent data leakage, the train-validation-test split was performed at the trial level rather than at the individual window level. A trial represents a single instance of QuaDRiGa simulation with a given parameters of receiver trajectory and channel parameters. Consequently, all histogram windows extracted from a given trial were assigned to only one subset.

\begin{figure}[!htbp]
    \centering
    \includegraphics[width=1\linewidth]{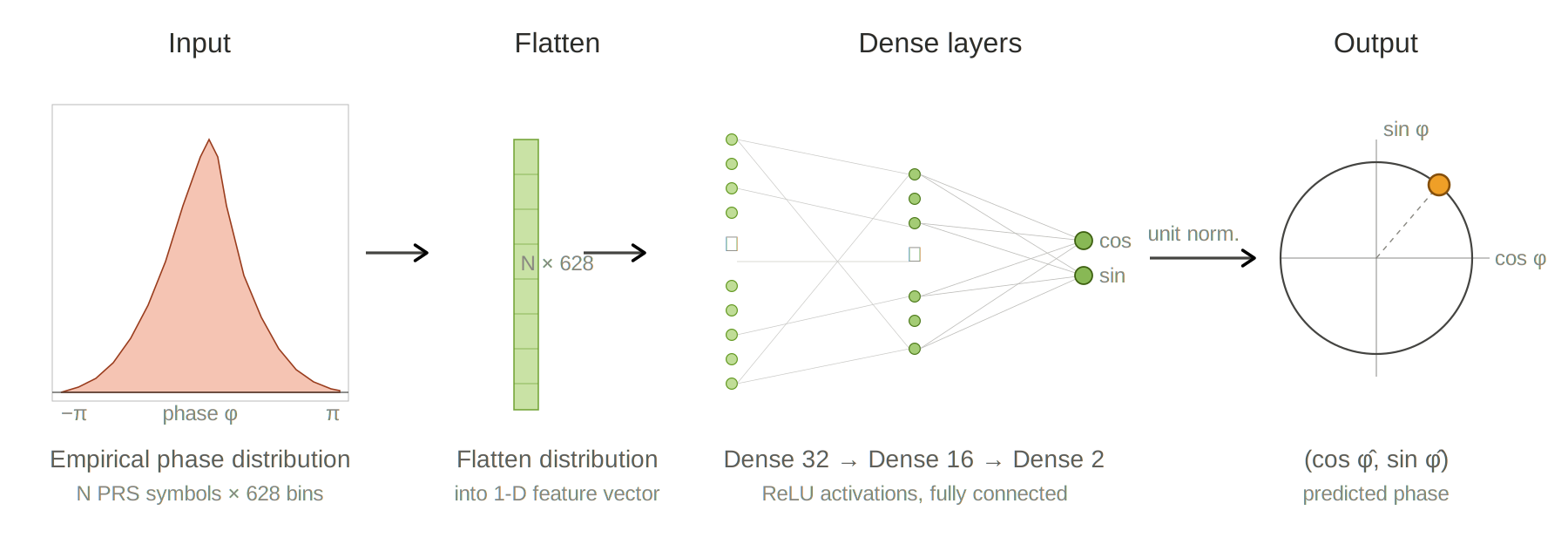} 
    \caption{Architecture of the FC phase estimator.}
    \label{fig:nn_arch_fc}
\end{figure}

\begin{figure}[!htbp]
    \centering
    \includegraphics[width=1\linewidth]{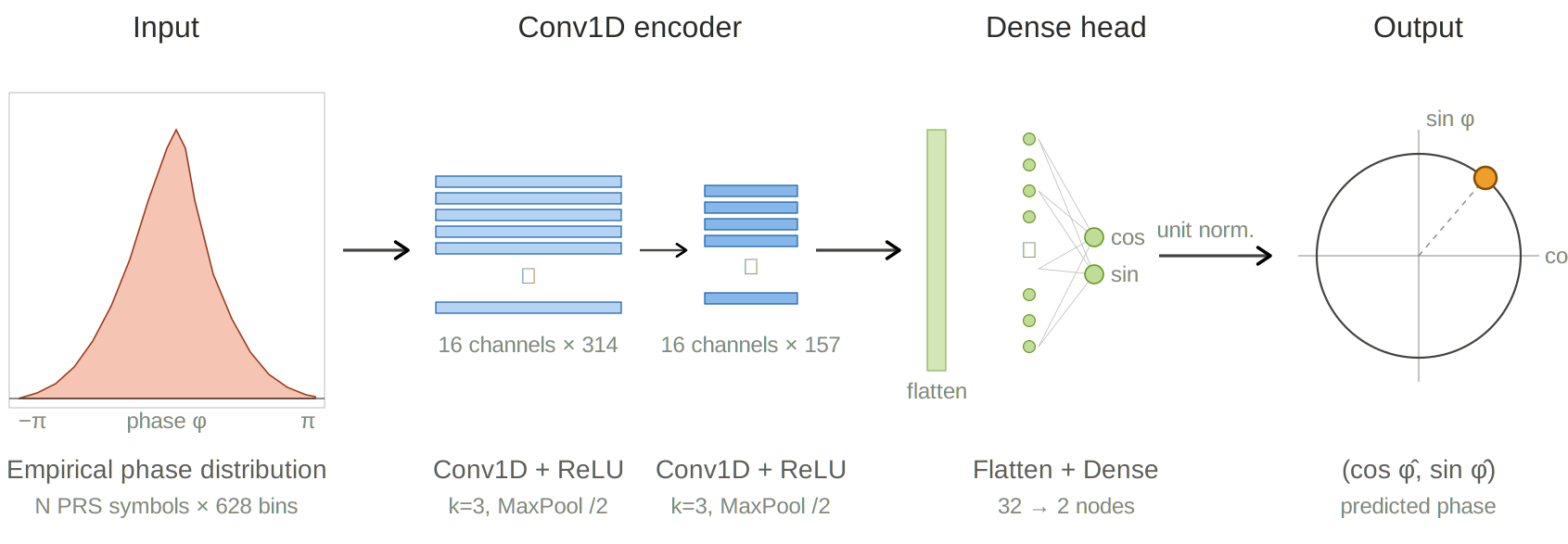} 
    \caption{Architecture of the 1D CNN phase estimator.}
    \label{fig:nn_arch_cnn}
\end{figure}

\begin{figure}[!htbp]
    \centering
    \includegraphics[width=1\linewidth]{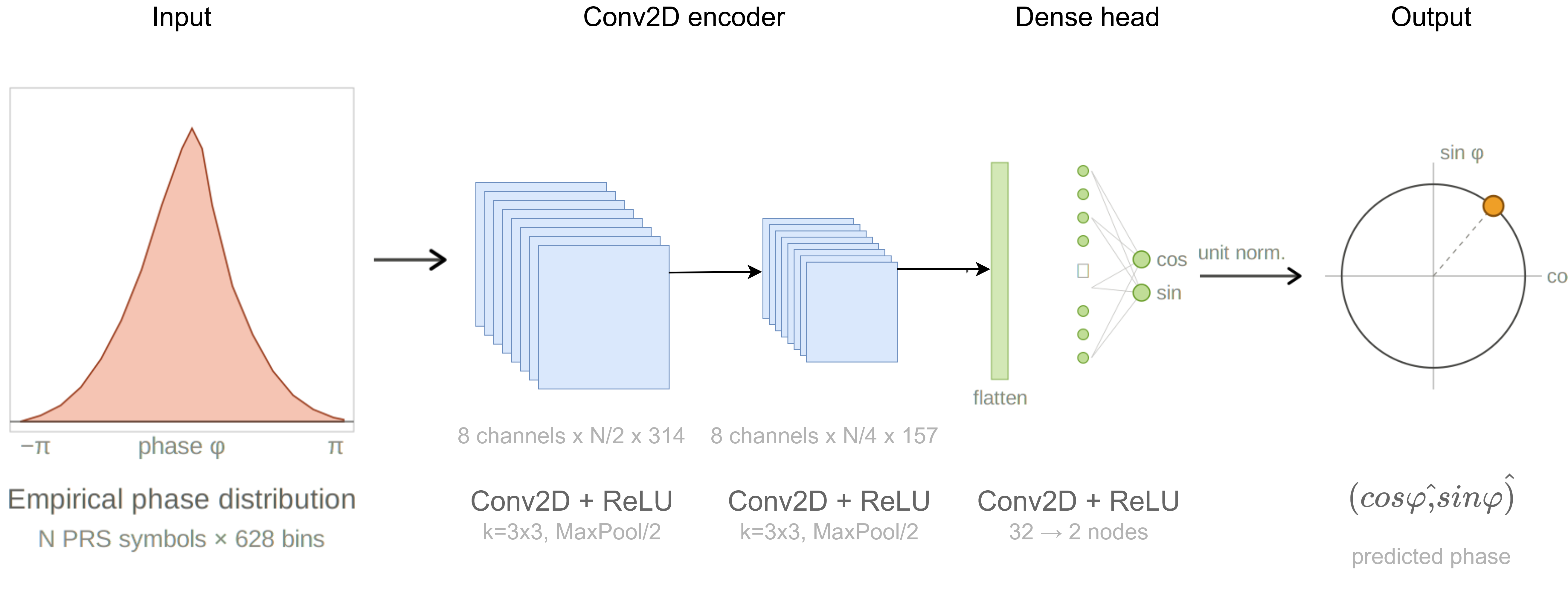} 
    \caption{\textcolor{black}{Architecture of the 2D CNN phase estimator.}}
    \label{fig:nn_arch_cnn}
\end{figure}


\textcolor{black}{The distinguishing feature of the proposed estimator lies in the input representation rather than in the depth of the network: the estimator operates on a normalised empirical histogram of inter-subcarrier phase differences defined on the circular domain. Within this representation, a dominant specular cluster appears as a compact, localised concentration of probability mass — precisely the type of structure that a convolutional operator acting along the bin axis is well suited to detect, and which is discarded by first-moment aggregation. Each architecture is designed with respect to this representation. The fully connected model provides a low-complexity reference that makes no locality assumptions. The 1D CNN exploits locality along the bin axis. The 2D CNN additionally exploits the PRS-symbol axis, and is therefore capable of identifying components that remain consistent across symbols — a signature characteristic of deterministic reflections rather than of noise.}

\section{Experimental Evaluation}
\label{sec:exp}

The proposed models are evaluated against the baseline circular mean estimator across both UMi LOS and InF LOS scenarios. \textcolor{black}{Performance is quantified using circular statistical metrics, including Mean Absolute Error (MAE), Root Mean Square Error (RMSE), and 95th percentile (P95).}

To evaluate the impact of temporal aggregation on estimator performance, multiple input configurations were considered for the neural networks. Input containing a single histogram was tested using the Fully Connected (FC) model. Subsequently, networks were trained using samples composed of consecutive PRS symbols, including 12-symbol and 60-symbol configurations, to assess the effect of increasing temporal context on estimation robustness. In addition, an alternative approach was investigated, where a single PRS symbol per slot was selected to incorporate temporal diversity while limiting the input dimensionality. Two configurations were evaluated in this case, using 5 and 10 PRS symbols collected across consecutive slots. 

\textcolor{black}{For each sample in the test dataset, three baseline estimates were computed. The first estimator is based on fitting a bimodal von Mises distribution, with the phase estimate taken as the center of the component having the highest weighted peak. The second estimator is the conventional circular mean. The third estimator is an amplitude-weighted circular mean, as introduced among the reference approaches in Section~\ref{sec:phase}, where the contribution of each phase observation is weighted by the product of the
amplitudes of the corresponding subcarrier pair. In test scenarios where consecutive symbols were used as input, the phase observations were flattened into a single vector before applying both estimators.} \textcolor{black}{The results for the UMi LOS and InF LOS scenario under Rx velocity of 30km/h with two different SNR groups are presented in Table~\ref{tab:results_umi_inf_30kmh}, while the corresponding results for the Rx velocity of 3 km/h are shown in Table~\ref{tab:results_umi_inf_3kmh}.}















{
\color{black}

\begin{table}[]
    \caption{Comparative performance of phase estimation methods for different sample sizes in InF LOS scenario and UMi LOS scenario under different SNR levels (30 km/h)}
    \centering
    \small

    \begin{tabular}{c|c|c|c|c|c|c|c|c|c|c|c|c|}
        \multirow{5}{*}{Estimator}
        & \multicolumn{12}{c|}{Scenarios} \\

        & \multicolumn{6}{c|}{InF}
        & \multicolumn{6}{c|}{UMi} \\

        & \multicolumn{3}{c|}{SNR}
        & \multicolumn{3}{c|}{SNR}
        & \multicolumn{3}{c|}{SNR}
        & \multicolumn{3}{c|}{SNR} \\

        & \multicolumn{3}{c|}{3--6}
        & \multicolumn{3}{c|}{10--12}
        & \multicolumn{3}{c|}{3--6}
        & \multicolumn{3}{c|}{10--12} \\

        & MAE & RMSE & P95
        & MAE & RMSE & P95
        & MAE & RMSE & P95
        & MAE & RMSE & P95 \\

        \textit{1 symbol } & & & & & & & & & & & &\\

         FC
         & {\small 0.121} & {\small 0.358} & {\small 0.439}
         & {\small 0.125} & {\small 0.350} & {\small 0.411}
         & {\small 0.034} & {\small 0.091} & {\small 0.156}
         & {\small 0.031} & {\small 0.098} & {\small 0.163} \\

         Circular mean
         & {\small 0.126} & {\small 0.381} & {\small 0.634}
         & {\small 0.120} & {\small 0.378} & {\small 0.624}
         & {\small 0.067} & {\small 0.186} & {\small 0.456}
         & {\small 0.057} & {\small 0.178} & {\small 0.425} \\

         W. circ. mean
         & {\small 0.152} & {\small 0.404} & {\small 0.730}
         & {\small 0.151} & {\small 0.404} & {\small 0.728}
         & {\small 0.106} & {\small 0.222} & {\small 0.568}
         & {\small 0.106} & {\small 0.221} & {\small 0.567} \\

         Bim. von Mises
         & {\small 0.198} & {\small 0.420} & {\small 0.934}
         & {\small 0.203} & {\small 0.420} & {\small 0.936}
         & {\small 0.128} & {\small 0.235} & {\small 0.588}
         & {\small 0.135} & {\small 0.237} & {\small 0.588} \\
         \hline
        
         \textit{12 symbols} & & & & & & & & & & & &\\

         FC
         & {\small 0.145} & {\small 0.398} & {\small 0.565}
         & {\small 0.143} & {\small 0.406} & {\small 0.502}
         & {\small 0.033} & {\small 0.081} & {\small 0.125}
         & {\small 0.031} & {\small 0.080} & {\small 0.142} \\

         1D CNN
         & {\small 0.155} & {\small 0.381} & {\small 0.567}
         & {\small 0.156} & {\small 0.372} & {\small 0.608}
         & {\small 0.030} & {\small 0.082} & {\small 0.122}
         & {\small 0.028} & {\small 0.083} & {\small 0.113} \\

         2D CNN
         & {\small 0.145} & {\small 0.370} & {\small 0.499}
         & {\small 0.168} & {\small 0.425} & {\small 0.502}
         & {\small 0.029} & {\small 0.083} & {\small 0.125}
         & {\small 0.030} & {\small 0.086} & {\small 0.153} \\

         Circular mean
         & {\small 0.124} & {\small 0.380} & {\small 0.629}
         & {\small 0.120} & {\small 0.378} & {\small 0.622}
         & {\small 0.066} & {\small 0.186} & {\small 0.455}
         & {\small 0.057} & {\small 0.178} & {\small 0.424} \\

         W. circ. mean
         & {\small 0.151} & {\small 0.403} & {\small 0.727}
         & {\small 0.151} & {\small 0.403} & {\small 0.722}
         & {\small 0.106} & {\small 0.221} & {\small 0.567}
         & {\small 0.106} & {\small 0.221} & {\small 0.567} \\

         Bim. von Mises
         & {\small 0.193} & {\small 0.416} & {\small 0.924}
         & {\small 0.200} & {\small 0.418} & {\small 0.932}
         & {\small 0.125} & {\small 0.232} & {\small 0.579}
         & {\small 0.134} & {\small 0.236} & {\small 0.587} \\
         \hline

         \textit{60 symbols} & & & & & & & & & & & &\\

         FC
         & {\small 0.148} & {\small 0.378} & {\small 0.572}
         & {\small 0.149} & {\small 0.403} & {\small 0.539}
         & {\small 0.042} & {\small 0.091} & {\small 0.203}
         & {\small 0.039} & {\small 0.079} & {\small 0.177} \\

         1D CNN
         & {\small 0.138} & {\small 0.369} & {\small 0.424}
         & {\small 0.132} & {\small 0.365} & {\small 0.380}
         & {\small 0.041} & {\small 0.099} & {\small 0.197}
         & {\small 0.033} & {\small 0.080} & {\small 0.144} \\

         2D CNN
         & {\small 0.169} & {\small 0.395} & {\small 0.607}
         & {\small 0.140} & {\small 0.388} & {\small 0.400}
         & {\small 0.039} & {\small 0.100} & {\small 0.185}
         & {\small 0.030} & {\small 0.077} & {\small 0.130} \\

         Circular mean
         & {\small 0.123} & {\small 0.377} & {\small 0.639}
         & {\small 0.118} & {\small 0.374} & {\small 0.629}
         & {\small 0.066} & {\small 0.184} & {\small 0.457}
         & {\small 0.057} & {\small 0.176} & {\small 0.431} \\

         W. circ. mean
         & {\small 0.149} & {\small 0.401} & {\small 0.733}
         & {\small 0.149} & {\small 0.401} & {\small 0.730}
         & {\small 0.106} & {\small 0.221} & {\small 0.568}
         & {\small 0.106} & {\small 0.221} & {\small 0.569} \\

         Bim. von Mises
         & {\small 0.191} & {\small 0.409} & {\small 0.919}
         & {\small 0.198} & {\small 0.415} & {\small 0.920}
         & {\small 0.124} & {\small 0.229} & {\small 0.579}
         & {\small 0.134} & {\small 0.235} & {\small 0.588} \\
         \hline

         \textit{5 symbols} & & & & & & & & & & & &\\

         FC
         & {\small 0.147} & {\small 0.388} & {\small 0.522}
         & {\small 0.140} & {\small 0.380} & {\small 0.531}
         & {\small 0.039} & {\small 0.090} & {\small 0.185}
         & {\small 0.041} & {\small 0.096} & {\small 0.200} \\

         1D CNN
         & {\small 0.135} & {\small 0.341} & {\small 0.458}
         & {\small 0.152} & {\small 0.380} & {\small 0.470}
         & {\small 0.040} & {\small 0.091} & {\small 0.165}
         & {\small 0.038} & {\small 0.087} & {\small 0.172} \\

         2D CNN
         & {\small 0.145} & {\small 0.346} & {\small 0.499}
         & {\small 0.158} & {\small 0.380} & {\small 0.558}
         & {\small 0.045} & {\small 0.105} & {\small 0.202}
         & {\small 0.036} & {\small 0.080} & {\small 0.155} \\

         Circular mean
         & {\small 0.123} & {\small 0.377} & {\small 0.618}
         & {\small 0.118} & {\small 0.374} & {\small 0.612}
         & {\small 0.066} & {\small 0.185} & {\small 0.454}
         & {\small 0.057} & {\small 0.177} & {\small 0.431} \\

         W. circ. mean
         & {\small 0.150} & {\small 0.402} & {\small 0.717}
         & {\small 0.150} & {\small 0.403} & {\small 0.720}
         & {\small 0.106} & {\small 0.221} & {\small 0.565}
         & {\small 0.106} & {\small 0.221} & {\small 0.562} \\

         Bim. von Mises
         & {\small 0.192} & {\small 0.415} & {\small 0.925}
         & {\small 0.198} & {\small 0.413} & {\small 0.931}
         & {\small 0.125} & {\small 0.231} & {\small 0.589}
         & {\small 0.133} & {\small 0.235} & {\small 0.590} \\
         \hline

         \textit{10 symbols} & & & & & & & & & & & &\\

         FC
         & {\small 0.143} & {\small 0.349} & {\small 0.501}
         & {\small 0.150} & {\small 0.384} & {\small 0.524}
         & {\small 0.045} & {\small 0.095} & {\small 0.200}
         & {\small 0.043} & {\small 0.093} & {\small 0.192} \\

         1D CNN
         & {\small 0.148} & {\small 0.359} & {\small 0.483}
         & {\small 0.131} & {\small 0.365} & {\small 0.432}
         & {\small 0.039} & {\small 0.092} & {\small 0.181}
         & {\small 0.036} & {\small 0.085} & {\small 0.146} \\

         2D CNN
         & {\small 0.132} & {\small 0.341} & {\small 0.409}
         & {\small 0.135} & {\small 0.353} & {\small 0.508}
         & {\small 0.044} & {\small 0.104} & {\small 0.176}
         & {\small 0.042} & {\small 0.095} & {\small 0.190} \\

         Circular mean
         & {\small 0.119} & {\small 0.370} & {\small 0.599}
         & {\small 0.114} & {\small 0.365} & {\small 0.586}
         & {\small 0.065} & {\small 0.181} & {\small 0.459}
         & {\small 0.056} & {\small 0.173} & {\small 0.425} \\

         W. circ. mean
         & {\small 0.147} & {\small 0.399} & {\small 0.707}
         & {\small 0.147} & {\small 0.400} & {\small 0.712}
         & {\small 0.106} & {\small 0.219} & {\small 0.559}
         & {\small 0.106} & {\small 0.220} & {\small 0.557} \\

         Bim. von Mises
         & {\small 0.187} & {\small 0.403} & {\small 0.894}
         & {\small 0.190} & {\small 0.405} & {\small 0.909}
         & {\small 0.124} & {\small 0.227} & {\small 0.577}
         & {\small 0.132} & {\small 0.231} & {\small 0.569} \\
         \hline
    \end{tabular}

    \label{tab:results_umi_inf_30kmh}
\end{table}
}













{
\color{black}
\begin{table}[]
    \caption{Comparative performance of phase estimation methods for different sample sizes in InF LOS scenario and UMi LOS scenario under different SNR levels (3 km/h)}
    \centering
    \small

    \begin{tabular}{c|c|c|c|c|c|c|c|c|c|c|c|c|}
        \multirow{5}{*}{Estimator}
        & \multicolumn{12}{c|}{Scenarios} \\

        & \multicolumn{6}{c|}{InF}
        & \multicolumn{6}{c|}{UMi} \\

        & \multicolumn{3}{c|}{SNR}
        & \multicolumn{3}{c|}{SNR}
        & \multicolumn{3}{c|}{SNR}
        & \multicolumn{3}{c|}{SNR} \\

        & \multicolumn{3}{c|}{3--6}
        & \multicolumn{3}{c|}{10--12}
        & \multicolumn{3}{c|}{3--6}
        & \multicolumn{3}{c|}{10--12} \\

        & MAE & RMSE & P95
        & MAE & RMSE & P95
        & MAE & RMSE & P95
        & MAE & RMSE & P95 \\

        \textit{1 symbol } & & & & & & & & & & & &\\
         FC
         & {\small 0.084} & {\small 0.264} & {\small 0.446}
         & {\small 0.091} & {\small 0.295} & {\small 0.507}
         & {\small 0.032} & {\small 0.180} & {\small 0.150}
         & {\small 0.034} & {\small 0.184} & {\small 0.093} \\

         Circular mean
         & {\small 0.109} & {\small 0.328} & {\small 0.620}
         & {\small 0.106} & {\small 0.330} & {\small 0.618}
         & {\small 0.115} & {\small 0.389} & {\small 1.043}
         & {\small 0.106} & {\small 0.387} & {\small 1.019} \\

         W. circ. mean
         & {\small 0.127} & {\small 0.331} & {\small 0.663}
         & {\small 0.126} & {\small 0.331} & {\small 0.663}
         & {\small 0.157} & {\small 0.402} & {\small 1.157}
         & {\small 0.157} & {\small 0.402} & {\small 1.162} \\

         Bim. von Mises
         & {\small 0.168} & {\small 0.366} & {\small 0.797}
         & {\small 0.180} & {\small 0.376} & {\small 0.830}
         & {\small 0.179} & {\small 0.403} & {\small 1.017}
         & {\small 0.191} & {\small 0.399} & {\small 0.995} \\
         \hline
         
         \textit{12 symbols } & & & & & & & & & & & &\\
         FC
         & {\small 0.101} & {\small 0.331} & {\small 0.449}
         & {\small 0.093} & {\small 0.269} & {\small 0.474}
         & {\small 0.032} & {\small 0.148} & {\small 0.068}
         & {\small 0.039} & {\small 0.182} & {\small 0.109}\\

         1D CNN
         & {\small 0.119} & {\small 0.320} & {\small 0.572}
         & {\small 0.091} & {\small 0.279} & {\small 0.476}
         & {\small 0.032} & {\small 0.144} & {\small 0.114}
         & {\small 0.041} & {\small 0.187} & {\small 0.121}\\

         2D CNN
         & {\small 0.075} & {\small 0.210} & {\small 0.390}
         & {\small 0.085} & {\small 0.227} & {\small 0.438}
         & {\small 0.028} & {\small 0.123} & {\small 0.105}
         & {\small 0.031} & {\small 0.141} & {\small 0.097}\\
         
         Circular mean
         & {\small 0.108} & {\small 0.328} & {\small 0.614}
         & {\small 0.106} & {\small 0.329} & {\small 0.616}
         & {\small 0.115} & {\small 0.388} & {\small 1.046}
         & {\small 0.106} & {\small 0.386} & {\small 1.017}\\

         W. circ. mean
         & {\small 0.126} & {\small 0.331} & {\small 0.662}
         & {\small 0.126} & {\small 0.331} & {\small 0.662}
         & {\small 0.157} & {\small 0.402} & {\small 1.164}
         & {\small 0.157} & {\small 0.402} & {\small 1.164}\\
         
         Bim. von Mises
         & {\small 0.165} & {\small 0.365} & {\small 0.808}
         & {\small 0.178} & {\small 0.373} & {\small 0.830}
         & {\small 0.177} & {\small 0.401} & {\small 1.001}
         & {\small 0.190} & {\small 0.399} & {\small 0.994}\\
         \hline
         
         \textit{60 symbols } & & & & & & & & & & & &\\
         FC
         & {\small 0.103} & {\small 0.291} & {\small 0.462}
         & {\small 0.078} & {\small 0.198} & {\small 0.395}
         & {\small 0.043} & {\small 0.162} & {\small 0.092}
         & {\small 0.051} & {\small 0.198} & {\small 0.136}\\

         1D CNN
         & {\small 0.104} & {\small 0.241} & {\small 0.497}
         & {\small 0.091} & {\small 0.241} & {\small 0.542}
         & {\small 0.041} & {\small 0.167} & {\small 0.075}
         & {\small 0.044} & {\small 0.197} & {\small 0.087}\\

         2D CNN
         & {\small 0.108} & {\small 0.231} & {\small 0.509}
         & {\small 0.105} & {\small 0.244} & {\small 0.496}
         & {\small 0.052} & {\small 0.182} & {\small 0.115}
         & {\small 0.062} & {\small 0.242} & {\small 0.171}\\
         
         Circular mean
         & {\small 0.108} & {\small 0.328} & {\small 0.618}
         & {\small 0.106} & {\small 0.329} & {\small 0.615}
         & {\small 0.115} & {\small 0.388} & {\small 1.062}
         & {\small 0.106} & {\small 0.386} & {\small 1.026}\\

         W. circ. mean
         & {\small 0.126} & {\small 0.331} & {\small 0.664}
         & {\small 0.126} & {\small 0.331} & {\small 0.664}
         & {\small 0.157} & {\small 0.402} & {\small 1.164}
         & {\small 0.157} & {\small 0.402} & {\small 1.168}\\
         
         Bim. von Mises
         & {\small 0.163} & {\small 0.363} & {\small 0.804}
         & {\small 0.174} & {\small 0.364} & {\small 0.797}
         & {\small 0.174} & {\small 0.399} & {\small 1.027}
         & {\small 0.191} & {\small 0.399} & {\small 0.946}\\
         \hline

         \textit{5 symbols } & & & & & & & & & & & &\\
         FC
         & {\small 0.091} & {\small 0.231} & {\small 0.455}
         & {\small 0.090} & {\small 0.245} & {\small 0.383}
         & {\small 0.038} & {\small 0.143} & {\small 0.119}
         & {\small 0.050} & {\small 0.219} & {\small 0.122}\\

         1D CNN
         & {\small 0.095} & {\small 0.227} & {\small 0.451}
         & {\small 0.088} & {\small 0.274} & {\small 0.377}
         & {\small 0.037} & {\small 0.133} & {\small 0.134}
         & {\small 0.050} & {\small 0.213} & {\small 0.107}\\

         2D CNN
         & {\small 0.093} & {\small 0.204} & {\small 0.522}
         & {\small 0.078} & {\small 0.198} & {\small 0.486}
         & {\small 0.045} & {\small 0.197} & {\small 0.093}
         & {\small 0.049} & {\small 0.202} & {\small 0.076}\\
         
         Circular mean
         & {\small 0.108} & {\small 0.328} & {\small 0.618}
         & {\small 0.105} & {\small 0.329} & {\small 0.615}
         & {\small 0.115} & {\small 0.389} & {\small 1.049}
         & {\small 0.106} & {\small 0.387} & {\small 1.025}\\

         W. circ. mean
         & {\small 0.126} & {\small 0.331} & {\small 0.663}
         & {\small 0.126} & {\small 0.331} & {\small 0.662}
         & {\small 0.157} & {\small 0.402} & {\small 1.164}
         & {\small 0.157} & {\small 0.402} & {\small 1.170}\\
         
         Bim. von Mises
         & {\small 0.165} & {\small 0.366} & {\small 0.799}
         & {\small 0.179} & {\small 0.377} & {\small 0.834}
         & {\small 0.177} & {\small 0.403} & {\small 1.025}
         & {\small 0.192} & {\small 0.400} & {\small 0.997}\\
         \hline

         \textit{10 symbols } & & & & & & & & & & & &\\
         FC
         & {\small 0.082} & {\small 0.191} & {\small 0.375}
         & {\small 0.072} & {\small 0.187} & {\small 0.242}
         & {\small 0.046} & {\small 0.163} & {\small 0.112}
         & {\small 0.044} & {\small 0.165} & {\small 0.111}\\

         1D CNN
         & {\small 0.103} & {\small 0.245} & {\small 0.448}
         & {\small 0.100} & {\small 0.245} & {\small 0.556}
         & {\small 0.049} & {\small 0.199} & {\small 0.117}
         & {\small 0.050} & {\small 0.223} & {\small 0.086}\\

         2D CNN
         & {\small 0.091} & {\small 0.227} & {\small 0.496}
         & {\small 0.097} & {\small 0.255} & {\small 0.492}
         & {\small 0.047} & {\small 0.133} & {\small 0.164}
         & {\small 0.055} & {\small 0.199} & {\small 0.108}\\
         
         Circular mean
         & {\small 0.108} & {\small 0.328} & {\small 0.611}
         & {\small 0.105} & {\small 0.328} & {\small 0.614}
         & {\small 0.115} & {\small 0.389} & {\small 1.057}
         & {\small 0.106} & {\small 0.387} & {\small 1.028}\\

         W. circ. mean
         & {\small 0.126} & {\small 0.331} & {\small 0.659}
         & {\small 0.126} & {\small 0.330} & {\small 0.658}
         & {\small 0.157} & {\small 0.403} & {\small 1.169}
         & {\small 0.157} & {\small 0.403} & {\small 1.171}\\
         
         Bim. von Mises
         & {\small 0.162} & {\small 0.358} & {\small 0.771}
         & {\small 0.178} & {\small 0.371} & {\small 0.816}
         & {\small 0.177} & {\small 0.402} & {\small 1.030}
         & {\small 0.190} & {\small 0.400} & {\small 1.009}\\
         \hline
    \end{tabular}

    \label{tab:results_umi_inf_3kmh}
\end{table}
}

\textcolor{black}{The results demonstrate that neural network-based estimators can outperform conventional baseline methods under specific scenarios. In the UMi LOS scenario, the learned estimators reduce both the average error and P95 relative to the circular mean and von Mises estimators. In the InF LOS scenario at 30 km/h, the circular mean achieves the lowest average error, whereas the learned estimators consistently reduce P95. Overall, the results indicate that the proposed approach improves the robustness of phase estimation. However, increasing the number of aggregated PRS symbols does not always lead to further improvements. Aggregating a large number of consecutive symbols may combine observations acquired under nearly identical channel conditions, increasing model complexity without introducing additional informative variability. In contrast, samples composed of fewer phase distributions but spanning a larger temporal separation can yield better performance. Moreover, the effect of the PRS aggregation window on the estimation accuracy is not consistent across the evaluated scenarios. The FC estimator operating on a single histogram derived from one PRS symbol achieves the lowest MAE and P95 in most evaluated scenarios, indicating that a single histogram already contains most of the information required for accurate phase estimation. Although temporal aggregation can improve robustness in selected cases, none of the evaluated aggregation strategies consistently outperforms the others across all considered propagation conditions.}

\textcolor{black}{It should also be noted that the phase estimator based on the fitted bimodal von Mises distribution generally achieves the highest errors across all evaluated scenarios. Although this mixture provides the best global fit to the phase-error distribution among the considered models, it is less effective when applied to sparsely represented empirical phase distributions. In such cases, the fitted dominant mode may be driven by a strong local concentration that does not correspond to the true phase of arrival, resulting in increased estimation errors.} 


\textcolor{black}{To provide a clearer interpretation of these results, the phase error is converted to a metric distance using the relationship $d = (\Delta\phi/2\pi)\lambda_v$. In this configuration, the achievable accuracy of any carrier-phase estimator is limited by the phase-to-distance conversion: for a given phase error, the resulting ranging error grows linearly with $\lambda_v$. Sub-metre accuracy would therefore require a considerably smaller virtual wavelength, which reintroduces the integer-ambiguity problem that the present configuration was specifically designed to avoid. The contribution of the proposed estimator should thus be understood as a reduction of phase-domain error --- particularly in its tail --- for a fixed $\lambda_v$, rather than as a demonstration of sub-metre ranging.}

For the InF LOS scenario, with a virtual wavelength $\lambda_v \approx 50$~m, the MAE of the circular mean estimator (0.328~rad) corresponds to a ranging error of approximately 2.61~m. The 1D CNN reduces the MAE to 0.232~rad, yielding an average error of 1.85~m, i.e., an improvement of 0.76~m.

The gain is even more pronounced for the P95 error, which reflects robustness to extreme multipath conditions. The baseline P95 of 2.098~rad corresponds to 16.70~m, while the CNN reduces it to 1.040~rad, or approximately 8.28~m. This nearly 8.4~m reduction highlights the ability of learning-based estimators to mitigate deterministic distortions in industrial propagation environments.

\section{Conclusion}
\label{sec:con}

This work investigated CP-based ranging under 3GPP-compliant propagation conditions, where deterministic multipath components lead to asymmetric and multimodal phase error distributions. The results confirm that classical circular averaging is insufficient in such environments, as the observed phase distributions cannot be reliably represented by simple parametric circular models.

The proposed learning-based estimators improve phase estimation accuracy by operating directly on empirical phase distributions. In the most demanding InF LOS scenario, the 1D CNN reduced the MAE from 0.328~rad to 0.232~rad, confirming that the structure of the observed phase data can be effectively exploited for estimation. \textcolor{black}{In most scenarios learning based estimators outperformed non parametric phase estimators}

The experiments also show that the temporal organization of the input data is important. Consecutive PRS symbols do not always provide additional information when the channel state remains highly correlated, whereas samples with larger temporal separation can improve performance as long as the receiver displacement remains small compared to the virtual wavelength ($\Delta d \ll \lambda_v$). \textcolor{black}{However, the effect of temporal aggregation on the estimation accuracy was not consistent across the evaluated scenarios, indicating that the relationship between the aggregation strategy and estimation performance requires further investigation.}

Finally, representing phase through $[\cos\phi,\sin\phi]$ enables stable learning in the circular domain by avoiding the discontinuity at the $2\pi$ boundary. More generally, the proposed approach illustrates how structured observations can be used directly for inference in estimation problems where physical propagation effects are difficult to capture with explicit statistical models.

\textcolor{black}{It is also important to note that the learning-based models were trained exclusively on channels generated using the QuaDRiGa channel simulator. Consequently, the models may have learned phase-error characteristics associated with the cluster generation mechanisms specific to this simulator. Although QuaDRiGa is a well-established implementation of the 3GPP TR 38.901 stochastic channel model and provides statistically representative channel realizations for system-level evaluation, it remains a simulation framework. Therefore, it is currently unknown how well the trained models will generalize to multipath propagation encountered in real measurement environments or when using alternative channel generators.} This also suggests a broader direction for future work: extending the proposed distribution-based inference framework toward system-agnostic phase estimation, where structured observations can be exploited across different propagation conditions, signal configurations, and wireless technologies without requiring a dedicated statistical model for each case.

\bibliographystyle{cas-model2-names}

\bibliography{cas-refs}

\end{document}